# Ultrafast Optomagnonics in Ferrimagnetic Multi-Sublattice Garnets

Andrzej Stupakiewicz[1*] and Takuya Satoh[2†]

[1] *Faculty of Physics, University of Bialystok, 15-245 Bialystok, Poland*

[2] *Department of Physics, Tokyo Institute of Technology, Meguro, Tokyo 152-8551, Japan*

This review discusses the ultrafast magnetization dynamics within the gigahertz to terahertz frequency range in ferrimagnetic rare-earth iron garnets with different substitutions. In these garnets, the roles of spin–orbit and exchange interactions have been detected using femtosecond laser pulses via the inverse Faraday effect. The all-optical control of spin-wave and Kaplan–Kittel exchange resonance modes in different frequency ranges is shown. Generation and localization of the electric field distribution inside the garnet through the metal-bound surface plasmon-polariton strongly enhance the amplitude of the exchange resonance modes. The exchange resonance mode in yttrium iron garnets was observed using circularly polarized Raman spectroscopy. The results of this study may be utilized in the development of a wide class of optomagnonic devices in the gigahertz to terahertz frequency range.

## 1. Introduction

Laser-induced magnetization dynamics is a flourishing area of research concerned with a deep understanding of previously unexplored and often strongly non-equilibrium phenomena, enabling the coherent control of magnetization on ultrashort time scales [1]. Using modern laser light sources, one can easily witness processes occurring on timescales down to femtoseconds. This field offers an unprecedented capability to probe light–matter interactions and manipulate magnetism at deep levels. Fundamental magnetic interactions, such as the exchange interaction responsible for magnetic ordering and the spin–orbit coupling responsible for magnetic anisotropy, have characteristic interaction times ranging from 1 ps to a few femtoseconds [2]. The practical applications include solving the most pressing issues related to modern technological developments, future generations of electronic devices, ultrafast processing of information, and development of spintronics and photonics.

Modern optical and magneto-optical methods offer new possibilities to study magnetization dynamics with spatial and temporal resolutions over a wide frequency range [3–6]. The laser-induced magnetization dynamics represent a fundamentally challenging physics, since they are typically strongly non-equilibrium, and the proper theoretical understanding is complicated by the multitude of couplings between the microscopic constituents of matter. Additionally, there is a fundamental distinction between the mechanisms allowing the manipulation of the magnetization by means of ultrashort laser pulses, and these phenomena can be classified into two types. First, the thermal effects, which are based on the ultrafast demagnetization with heating of the medium leading to the change in its magnetic properties, observed usually in metallic systems [7,8]. Second, non-thermal effects, which operate without heating the medium, are usually observed in the form of the inverse Faraday effect (IFE) [9,10] and the photomagnetic effect [11–13] in magnetic dielectrics, such as iron garnets with different dopants. The non-thermal effect are the most intriguing due to the dissipation-free mechanism of magnetization dynamics and all-optical switching of magnetization with lowest heat load. The non-thermal photomagnetic effect is based on absorption of photons during optical excitation of the strongly anisotropic ions using linear polarization of light. Such excitation allows one to generate an effective field of photo-induced magnetic anisotropy, which in general can have lifetime much longer than the laser pulse width itself and magnitude comparable to the intrinsic anisoropy; it is sufficient to trigger angle precession of the magnetization. On the contrary, the IFE does not require the absorption of the pump light in the materials. The lifetime of the effective magnetic field arising from the IFE is only as long as the pulse width, which is typically couple of tens to hundred femtoseconds. This fact allows the use of IFE as a mechanism of laser-induced control of magnetization for fundamental research and applications.

The first explanation of all-optical helicity-dependent magnetization switching [14] was based on the IFE using femtosecond circularly polarized laser pulses [15]. However, further investigations showed that the helicity dependence of all-optical switching, previously regarded as the confirmation of the non-thermal origin of effects, occurs due to circular dichroism, resulting in different heating efficiencies [16]. Difficulties with manipulation of magnetization in metals by IFE are associated with high absorption in the visible and near-infrared regions of spectra and as a result, weak IFE amplitude. However, the idea of using an impulsive magnetic field created by laser pulses for non-thermal manipulation of magnetization by IFE seems to be promising for highly transparent and magneto-optically active materials, such as ferrimagnetic dielectrics (e.g., garnets).

The IFE is the creation of an effective magnetic field in magnetic media during the action of a circularly polarized laser pulse. This effect was originally predicted by Faraday in the second part of his work concerning the interaction between light and magnetization [17]. In later theoretical works [18,19], it was shown that circularly polarized light can induce the "DC" magnetic field along the propagation direction of the incident light:

$$\mathbf{H}(0) = a[\mathbf{E}(\omega) \times \mathbf{E}^*(\omega)], \qquad (1)$$

where $a$ is the magneto-optical constant. The same constant $a$ is used in the direct Faraday effect and indicates magneto-optical coupling. Thus, the process of DC magnetic field generation via photons is called the IFE, which underlines the same magneto-optical constant presented in Eq. (1). The microscopic mechanism of the IFE was explained using impulsive stimulated Raman scattering [20,21]. This mechanism is based on a laser pulse stimulation of an optical transition from the ground state to a virtual excited state, which is split due to the spin–orbit coupling. Then the same laser pulse stimulates the transition back to the ground state, in which the magnetic state is changed.

The present review consists of five parts: (1) Introduction, (2) Experimental details describing the magnetic ordering in rare earth (RE) iron garnets and pump-probe magneto-optical method to laser-induced magnetization dynamics; (3) Laser-induced wide frequency range of magnetization dynamics in iron garnets, presenting the results of the excitation of precession of magnetization at different frequency ranges corresponding to the spin-wave and exchange resonance modes in RE iron garnets by IFE; (4) Generation and localization of magnetization precession under excitation of the exchange resonance mode in magneto-plasmonic Au/garnets, dedicated to the amplification of amplitude of magnetization precession with a nanoscale localization in depth of the garnet during excitation of exchange resonance mode via surface plasmon-polariton (SPP) resonance in Au/garnet structure; (5) Exchange resonance mode in yttrium iron garnet (YIG), devoted to study of the exchange resonance mode in YIG using circularly-polarized Raman spectroscopy.

## 2. Experimental Details
*2.1 RE iron garnets*

The magnetic order in a pure YIG ($Y_3Fe_5O_{12}$) is attributed to the localized spins of its magnetic ions and their super-exchange interaction with the surrounding oxygen ions. Due to the spatial overlap of the electron wave functions of the neighboring ions, the oxygen's $2p$ states mediate the interaction between $3d^5$ states of the iron ions close by [22]. The super-exchange of the overlapping orbitals brings about an extra energy term, whose nature depends on the relative arrangement of the electron's spins, the strongest being the super-exchange interaction between inequivalent tetrahedral (d) and octahedral [a] sites [Fig. 1(a)]. As the $Fe^{3+}$ are arranged in two separate tetra- and octahedral sublattices and the spins within each sublattice are ferromagnetically ordered, one can consider the sublattices to have own magnetizations (of various magnitudes). However, the iron sublattices themselves are coupled antiferromagnetically and the total magnetization is a sum of each sublattice contribution, producing a ferrimagnetic ordering. Various possibilities in garnet substitution, especially by magnetic RE ions in dodecahedral {c} sublattice provide the opportunity to vary properties over a wide range.

Generally, the $f$ shell can have more electrons, and in some cases, the RE ions produce larger magnetic moments than transition metal (TM) ions. However, because of their larger ionic radii and consequently longer RE-RE distances, their coupling is weaker than that of TM ions. This leads to differences in the temperature dependence of magnetization of sublattices. Doping with ions having large spin–orbit splitting parameter into dodecahedral sites can produce an increase in the Faraday rotation angle. Generally, the Faraday rotation in Bi-substituted RE iron garnets is strongly dependent on the wavelength in the visible range [23]. An enhanced Faraday rotation is usually accompanied by a large absorption [24,25]. In particular, a giant Faraday rotation is caused by an increase in the spin–orbit splitting in the excited states because of the formation of hybrid molecular orbitals between the $3d$ orbital in $Fe^{3+}$ and the $2p$ orbital in $O^{2-}$, mixed with the $6p$ orbital in $Bi^{3+}$, which has a large spin–orbit interaction parameter. In such materials, the magnetic moment of the RE sublattice originates from $4f$ localized electrons, in contrast to the iron sublattice, where the magnetism occurs by means of $3d$ delocalized conduction electrons. Consequently, the RE sublattice is polarized by the iron sublattice, and the magnetic moments become antiparallel through exchange interactions. One can expect that the exchange interaction between iron sublattices will be stronger than the exchange interaction between iron and RE (e.g., Gd and Yb ions) sublattices. We note here that the exchange between RE sublattices will be negligible in comparison to the above.

In our experiments, we used 380-μm-thick $Gd_{4/3}Yb_{2/3}BiFe_5O_{12}$ (GdYb-BIG), 200-μm-thick $Gd_2BiFe_5O_{12}$ (Gd-BIG), and 160-μm-thick YIG single crystals [26–28]. In this review, we also consider 7-μm-thick $Lu_{1.69}Y_{0.65}Bi_{0.66}Fe_{3.85}Ga_{1.15}O_{12}$ (LuIG) garnet grown on GGG substrate [29]. In these garnets, Gd, Yb, Lu, Y, and Bi ions are surrounded by eight oxygen ions in a dodecahedral sublattice {c}.

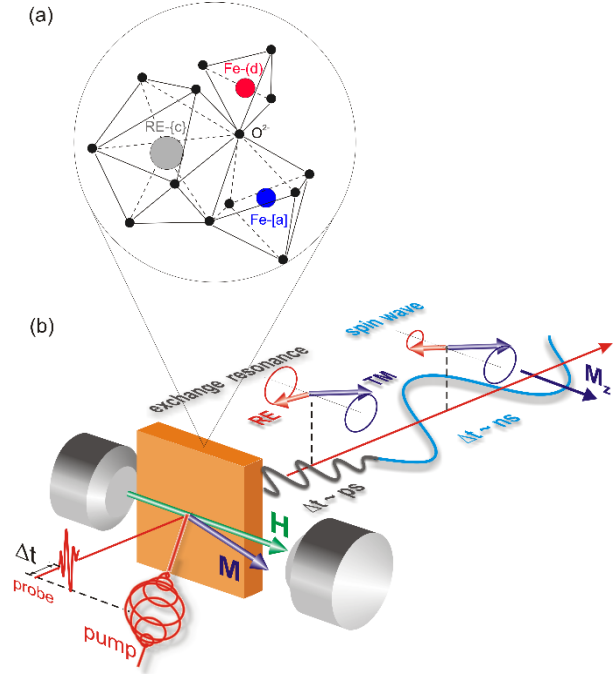

Fig. 1. (Color online) (a) Garnet lattice with octahedral-[a], dodecahedral-{c} and tetrahedral-(d) sites. (b) Visualization sketch of the magnetization precession and schematic illustration of the samples with experimental geometry.

*2.2 Time-resolved magneto-optical method*

As shown in Fig. 1(b), time-resolved pump-probe measurements were performed in a transmission geometry to investigate the ultrafast magnetization dynamics in ferrimagnetic RE iron garnets [26,27]. Pump pulses with a duration of 50 fs from an amplifier at a 500 Hz repetition rate were directed at an angle of incidence of approximately a few degrees from the sample normal. The probe laser pulses at a repetition rate of 1 kHz were incident normal to the sample. In order to keep off influence of most intense pump pulse on detector, the propagation direction of the pump pulse is generally not identical to that of the probe pulse. The wavelength of the pump beam was adjustable by an optical parametric amplifier in the near-infrared range of 1000–1500 nm, whereas the probe wavelength was fixed at 800 nm. The pump beam with an energy density of ~40 mJ/cm$^2$ was focused to a spot approximately 100 μm in diameter, while the 50 times weaker probe beam on the sample was twice as small in diameter. The delay time $\Delta t$ between the pump and probe pulses can be adjusted up to 2.5 ns. The polarization of the pump pulse was adjusted between circular polarization with different helicities. The polarization plane of the probe beam was linear. An external in-plane magnetic field $H_\parallel$ was applied to the sample. We measured the time-resolved Faraday rotation angle $\theta_F$ of the probe as a function of the delay time $\Delta t$. The angle of the Faraday rotation $\theta_F$ is proportional to the out-of-plane component of magnetization $M_x$ in a sample.

## 3. Laser-Induced Wide Frequency Range of Magnetization Dynamics in Iron Garnets

Here, we will consider the results of the non-thermal laser-induced ultrafast magnetization precession in Lu-iron garnet [29], and our results obtained for Gd- [27], and GdYb- [26] iron garnets. The magnetization precession in the picosecond and nanosecond time scales was optically excited via the IFE using the circular polarization of pump pulses. Two types of magnetization precession with different time scales were observed [Fig. 1(b)]: a low-frequency (ferromagnetic resonance (FMR) and spin-wave) mode with a frequency in the gigahertz (GHz) range for a duration of approximately 2.5 ns, and a high-frequency (HF) (exchange resonance) mode with a frequency in the sub-terahertz (sub-THz) range, which was present immediately after the initiation of the magnetization precession and lasted for approximately 30 ps.

*3.1 Magnetization precession in ferrimagnets*

The collective oscillations of excited spins in magnetic materials, such as the magnetization precession or spin waves, can be used to extract the information about the essential magnetic properties and phenomena. A clear understanding of the oscillations in the ferrimagnet can be obtained from the analysis of the equation of motion of each sublattice under an external magnetic field. Generally, the number of possible self-modes of oscillation is equal to the number of sublattices, which is in turn related to the number of different magnetic ions. The simplest case is a ferrimagnet with two sublattices. Neglecting the anisotropy field, demagnetizing field, and damping, the equations of motion for the two-sublattice systems $\mathbf{M_1}$ and $\mathbf{M_2}$ under the external magnetic field $\mathbf{H}$ are

$$\frac{d\mathbf{M_1}}{dt} = -\gamma_1[\mathbf{M_1} \times (\mathbf{H_1} + \mathbf{H})],$$
$$\frac{d\mathbf{M_2}}{dt} = -\gamma_2[\mathbf{M_2} \times (\mathbf{H_2} + \mathbf{H})].$$

(2)

The exchange fields are $\mathbf{H_1} = -\lambda \mathbf{M_2}$ and $\mathbf{H_2} = -\lambda \mathbf{M_1}$, where $\lambda$ is the molecular field coefficient between the two sublattices. $\gamma_1, \gamma_2 (> 0)$ are the gyromagnetic ratio. We set $\mathbf{M_1} = (m_1^x e^{i\omega t}, m_1^y e^{i\omega t}, M_1)$, $\mathbf{M_2} = (m_2^x e^{i\omega t}, m_2^y e^{i\omega t}, -M_2)$, and $\mathbf{H} = (0, 0, H_\parallel)$ and solve the equations of motion for $\mathbf{M_1}$ and $\mathbf{M_2}$. Here, $M_1 > M_2 > 0$ and $H_\parallel > 0$. Two solutions of the secular equation are possible, representing two types of oscillations [30,31]: FMR with a positive frequency $\omega_{\text{FMR}} = \gamma_{\text{eff}} H_\parallel$, and exchange resonance with a negative frequency

$$\omega_{\text{ex}} = -[\lambda(\gamma_2 M_1 - \gamma_1 M_2) - \gamma'_{\text{eff}} H_\parallel]. \tag{3}$$

Here, we also assume that the molecular fields $\mathbf{H_1}$ and $\mathbf{H_2}$ are much stronger than the external field $\mathbf{H}$. Further, assuming a thin film with the thickness direction parallel to *x* axis, uniaxial anisotropy field $H_A$ along *x* axis, demagnetizing field (demagnetization factor $\mathbf{N}$), and in-plane magnetic field $H_\parallel (> H_A)$ along *z* axis, the FMR frequency can be obtained by the Kittel formula [32]

$$\omega_{\text{FMR}} = \gamma_{\text{eff}} \sqrt{[4\pi(M_1 - M_2)(N_x - N_z) + H_\parallel - H_A] H_\parallel}. \tag{4}$$

For zero-orbital contribution to the magnetic moment of the ion, the gyromagnetic ratio for each sublattice can be the same: $\gamma_1 = \gamma_2 = \gamma$ and $\gamma_{\text{eff}} = \gamma'_{\text{eff}} = \gamma$. These types of precession are illustrated in Fig. 2. The parameter $\lambda_{\text{dc}}$ responsible for the exchange resonance mode is a molecular-field parameter between (d) and {c} [22]. $\lambda_{\text{dc}}$ is typically higher than $\lambda_{\text{ac}}$, which implies a stronger interaction. A positive $\lambda_{dc}$ corresponds to an antiferromagnetic interaction. Thus, the exchange resonance modes have an antiferromagnetic nature.

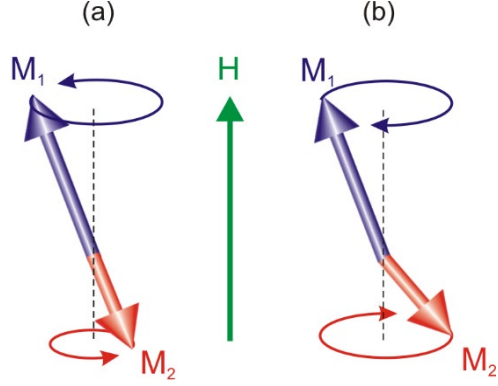

Fig. 2. (Color online) Schematic of (a) FMR (a) and (b) exchange resonance modes for a two-sublattice ferrimagnet.

3.2 *Spin-wave mode excitation via the IFE*

In order to determine the mechanism and characteristics of the laser-induced magnetization precession, we carried out measurements as functions of the pump polarization and the static applied magnetic field. Such measurements allow one to determine the excited magnetization precession mode, such as that under the change in frequency of the precession. The Bi-substituted iron garnets have the giant static Faraday rotation. Thus, all experimental data, which are presented below, were obtained using the GdYb-BIG sample at room temperature. However, the behaviors of the main intrinsic parameters of magnetization dynamics in the Gd-BIG sample at room temperature were similar. Figure 3 shows the experimental geometry and the magnetization precession as a time-resolved Faraday rotation angle for left-handed ($\sigma^+$) and right-handed ($\sigma^-$) circularly polarized pump pulses with in-plane external magnetic field $H_\parallel = 1.2$ kOe.

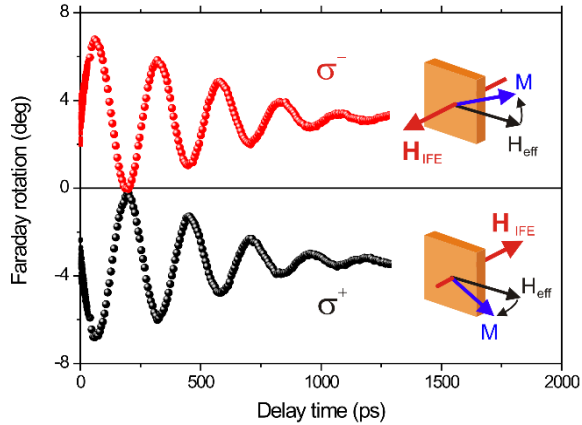

Fig. 3. (Color online) Time-resolved Faraday rotation for different helicities of the circular polarizations of the pump beam in an applied magnetic field $H_\parallel = 1.2$ kOe at room temperature for the GdYb-BIG sample. Insets show the graphical illustration of the excitation of magnetization precession by effective IFE field $H_{IFE}$.

Figure 4(a) shows the linear dependence of amplitude of magnetization precession for different pump laser pulses. Such behavior is typical for excitation of magnetization dynamics via IFE because this effect does not require photon absorption during laser excitation. For significantly higher pulse energies, the effect saturates. The dependence of the magnetization precession amplitude, which was extracted from time-traces measurements, as a function of a quarter-wave plate angle (compensator polarization) for the pump polarization is shown in Fig. 4(b). The magnetization precessions with opposite phases were triggered by $\sigma^+$ and $\sigma^-$ polarizations. Generally, the opposite phase precession of magnetization for different pump polarization (especially for different helicity of circular polarization) in ferrimagnets is the fingerprint of the non-thermal excitation. This result is consistent with a laser-induced IFE magnetic field $H_{IFE}$ along the direction perpendicular to the sample plane during a femtosecond laser pulse excitation in an iron garnet [9]. In this case, the magnetization vector precesses around the effective magnetic field ($H_{eff}$) as a result of the action of different contributions: the external in-plane magnetic field and magnetic anisotropy field (inset in Fig. 3).

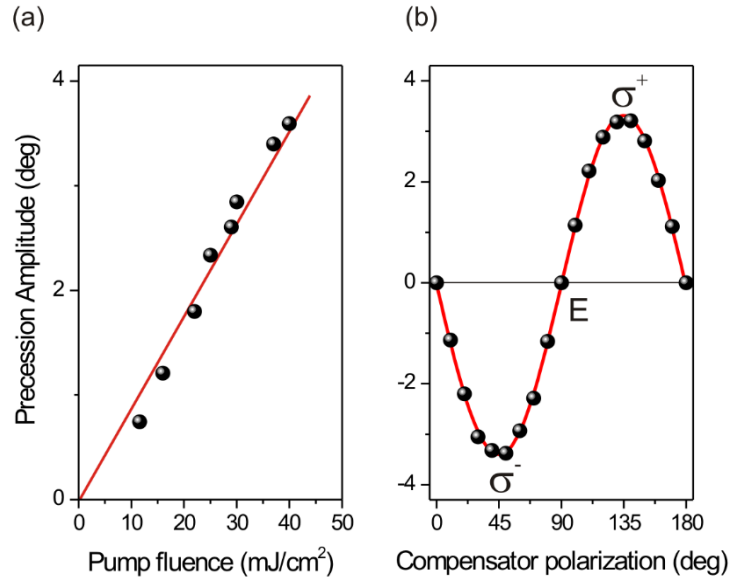

Fig. 4. (Color online) Dependence of the amplitude of magnetization precession as a function of (a) pump fluence and (b) compensator pump polarization (45° - left-handed cirlular polarization $\sigma^-$, 135°- left-handed cirlular polarization $\sigma^+$, and 90° - linear polarization E) at $H_\parallel = 1.2$ kOe in the GdYb-BIG sample.

In order to interpret the nature of the magnetization precession with respect to FMR mode, time-resolved measurements of magnetization dynamics have been performed with different amplitudes of external magnetic field larger than value of the magnetic anisotropy field $H_A$ = 620 Oe. The transient Faraday rotation as a function of delay time with various $H_\parallel$ are shown in Fig. 5. Behavior of these curves suggest the beating effect which can be determined using the analysis of spectrum frequencies of the magnetization precession. For this, we have performed the fast Fourier transform (FFT) which shows the spectrum of distinct frequencies of the magnetization precession (see the inset in Fig. 5).

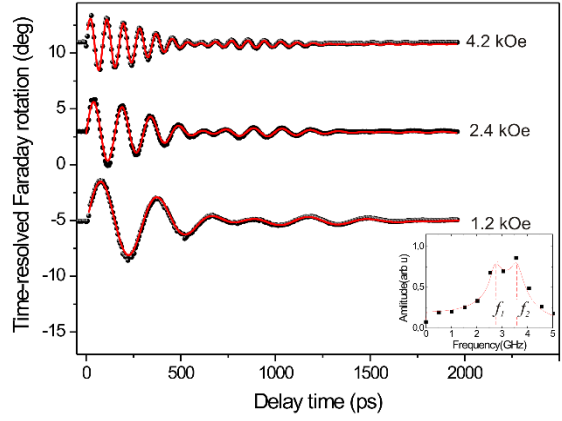

Fig. 5. (Color online) Time-resolved Faraday rotation as a function of the delay time for different amplitudes of external magnetic field $H_\parallel$. The red solid lines are fitted using Eq. (5). The inset is the FFT spectrum for $H_\parallel = 1.2$ kOe.

As presented in Ref. 10 for a similar garnet, the excitation and propagation of spin-wave modes using IFE has been shown. In such case, the spin-wave excitation in this garnet can induce a spatial distribution of the magnetization orientation within the spot of the pump beam, thus leading to a beating precession process. The two frequency components $f_1$ and $f_2$ originated from the backward-volume magnetostatic waves (BVMSWs) with wavevectors **k**. Thus, the results of laser-induced magnetization precession can be described as following function [10]:

$$\theta_\text{F}(\Delta t) = A \int d\mathbf{k} h(\mathbf{k}) \sin(\mathbf{k} \cdot \mathbf{r} - 2\pi f(\mathbf{k})\Delta t)\exp(-2\alpha\pi f(\mathbf{k})\Delta t), \qquad (5)$$

where $A$ is the amplitude and $h(\mathbf{k})$ is the FFT of the intensity distribution at the pump spot, and $\alpha$ is Gilbert damping. The results of the numerical calculations using Eq. (5) confirmed the results of the experiments and the FFT spectrum (inset in Fig. 5). In addition, experimental time-traces datasets were fitted using Eq. (5) and are shown in Fig. 5 (red solid lines). With increasing external magnetic field, significant changes were observed in the two frequencies $f_1$ and $f_2$ over the entire range of applied fields as shown in Fig. 6(a). In order to verify BVMSW modes, the dispersion dependence for frequency $f(\mathbf{k})$ have been calculated [Fig. 6(b)]. In addition, as shown, BVMSW mode can exist for **k** perpendicular to the in-plane orientation of magnetic field $H_\parallel$ [33].

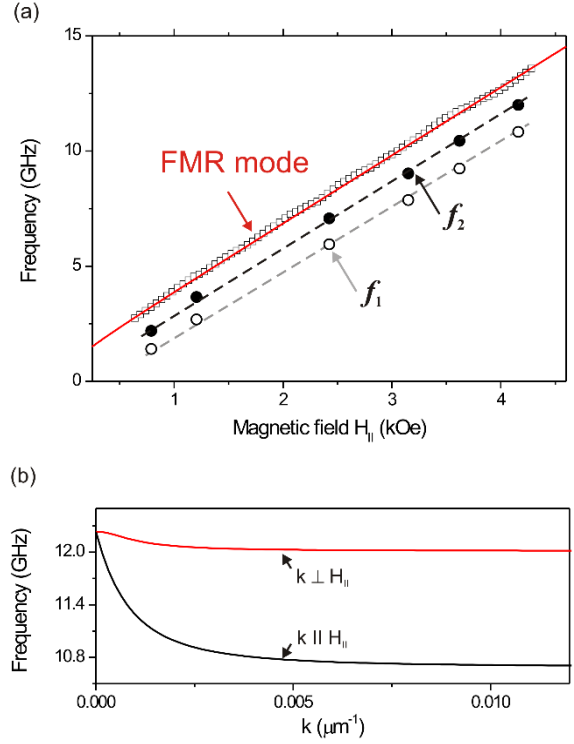

Fig. 6. (Color online) (a) Experimental data for the frequencies corresponding to spin-wave modes (circle dots) and FMR data (square dots) as a function of the external magnetic field. Red solid line represents the ferromagnetic resonance frequency results from the calculations using the Kittel formula Eq. (4). (b) Dispersion spectra of the spin-wave mode at $H_\parallel = 4.2$ kOe. Here, $k$ is defined as the inverse of the wavelength.

To compare the obtained spin-wave modes, we measured and calculated the frequency of FMR mode with the Kittel formula using Eq. (4), using $H_A = 620$ Oe and $4\pi M_S = 4\pi(M_{Fe} - M_{RE}) = 1140$ G. The magnetization of RE sublattices $M_{RE}$ was defined by the magnetization of Gd sublattice due to the zero contribution of the magnetization Yb ions at room temperature. In Fig. 6(a), it is clearly seen that the frequencies of laser-induced magnetization precession are lower than the measured and calculated FMR frequency [Fig. 6(a)].

*3.3 Laser-induced exchange resonance between iron sublattices in garnets*

Based on the Néel theory of ferrimagnetism, antiparallelly oriented non-equal magnetic moments reside in the exchange field created by themselves. In this case, the precession of sublattice magnetizations in the exchange field is the so-called Kaplan–Kittel exchange resonance [30]. Typically, in comparison with the FMR, a high value of the exchange field leads to a high frequency for such oscillations, which corresponds to the far-infrared (FIR) range. The FIR absorption and transmission spectra were measured for a number of ferrimagnets [34,35], and the exchange resonance frequency was determined. Nevertheless, there is poor information on the optical excitation of the exchange resonance in the visible and NIR regions of the spectrum.

The possibility of the excitation of ultrafast magnetization precession using the IFE without heating in LuIG has been reported [29]. In this case, the exchange resonance mode was excited between iron sublattices (tetrahedral and octahedral) $\mathbf{M}_d$ and $\mathbf{M}_a$ in LuIG with a frequency range of 400–700 GHz.

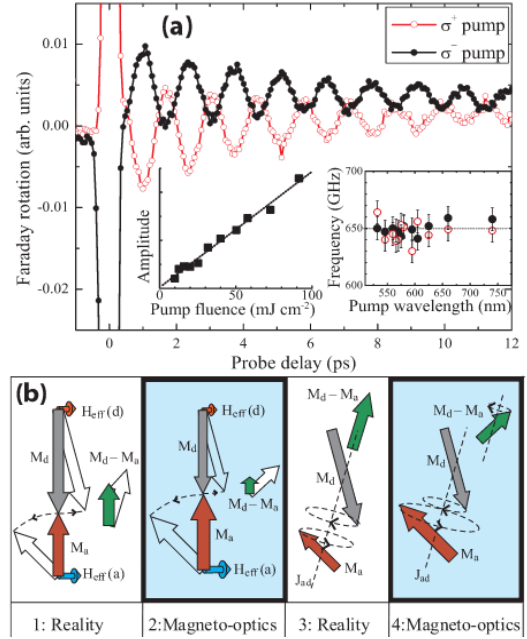

Fig. 7 (Color online) (a) Exchange resonance mode excited by $\sigma^+$ and $\sigma^-$ circularly polarized pump light at room temperature. The inset graphs show the amplitude dependence on the pump fluence and the spectral dependence of this mode. (b) Schematic illustration of the excitation of the exchange resonance. The figure is reproduced with permission from Ref. 29 (@2010 American Physical Society).

The nature of this mode is related to the canting of the two sublattices away from their mutual exchange interactions. The exchange resonance mode shows a 180°-phase shift between the excitations with $\sigma^+$ and $\sigma^-$ circularly polarized light [Fig. 7(a)]. No variations were observed in the amplitude or frequency of changes in a sufficiently small external field up to 3 kOe.

In this case, femtosecond pump pulses with circularly polarized light were used to excite the magnetic-dipole forbidden exchange resonance between the magnetizations of two iron sublattices via the IFE. The tetrahedral and octahedral iron sublattices experience inequivalent optically induced effective magnetic fields because of the different magneto-optical susceptibilities, resulting in a canting between magnetizations of the sublattices [Fig. 7(b)]. In Ref. 29, the contribution of dodecahedral ions was negligible, which play an important role, especially in Gd-substituted iron garnets.

*3.4 Laser-induced exchange resonance between RE and iron sublattices in garnets*

We experimentally demonstrated the laser-induced precessional motion of the magnetization with a frequency of approximately 410 GHz in GdYb-BIG at room temperature via the IFE [26]. In this HF mode, we observed a precession of oppotiste phases with respect to the pump helicity [Fig. 8(a)]. Figure 8(b) shows the precession dependencies for different amplitudes of magnetic field. In addition, the spin-wave mode was excited along with the HF mode. In a duration of approximately 20 ps, we observed the simultaneous relaxation of the HF mode and the excitation of the spin-wave mode at a frequency of a few GHz.

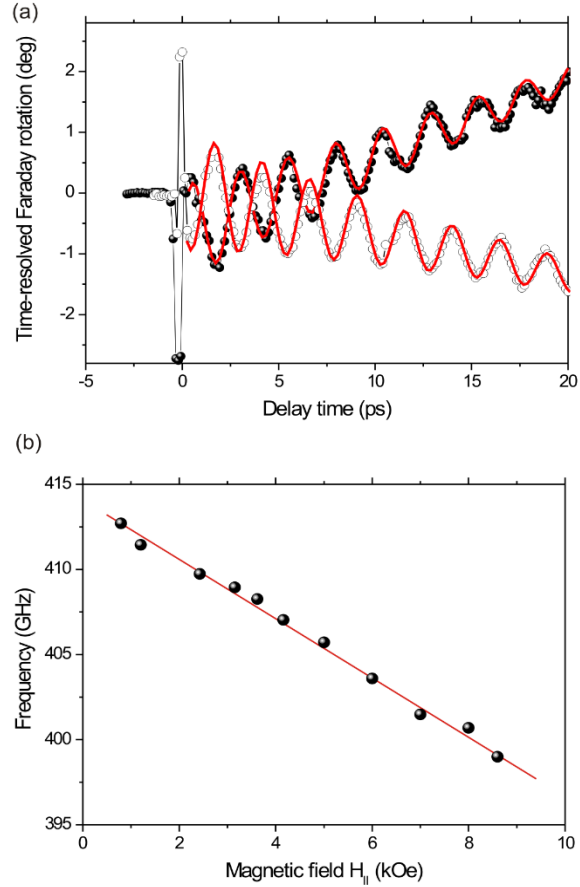

Fig. 8. (Color online) (a) Time-resolved Faraday rotation as a function of delay time for an external magnetic field of 1.2 kOe induced by σ⁺ (open points) and σ⁻ (full points) circularly polarized pump light at room temperature in the GdYb-BIG sample. (b) Exchange resonance frequency as a function of the external magnetic field measured at room temperature.

The frequency of the HF mode [Fig. 8(b)] linearly decreased with increasing intensity of the external magnetic field. Such dependence agrees with Eq. (3); however, the maximum achievable change in frequency was only 3.8%. The value of slope 1.74 MHz/Oe is lower than that for free electron motion, i.e., $\gamma = 2.8$ MHz/Oe. This dynamic behavior is a consequence of the exchange resonance between the magnetic ions in the RE and iron sublattices $\mathbf{M}_{RE}$ and $\mathbf{M}_{Fe}$. In this case, the frequency of exchange resonance is lower than the frequency resulting from exchange resonance between the iron sublattices (650 GHz at room temperature [29]). In contrast to our results for GdYb-BIG garnet, in LuIG garnet, changes were not observed in the exchange frequency for changes in the external field up to 3 kOe [29]. In RE garnets with non-zero contribution to the magnetization from the dodecahedral sites at room temperature, the exchange field between the iron and RE sublattices is weaker than that between the octahedral and tetrahedral iron sublattices. Therefore, influence of external magnetic field in relation to exchange resonance between RE and iron sublattices will be weaker than in case of the spin-wave or FMR modes [Fig. 6(a)].

*3.5 Temperature dependence of the exchange resonance mode*

Because of the high value of the exchange field, the dependence on the external magnetic field is not strong. Hence, the variation in temperature seems to be a suitable way to confirm the excitation of such modes as a consequence of the temperature dependence of the exchange field.

To confirm the prediction related to the excitation of the exchange resonance mode, temperature experiments were performed. Figure 9 shows the time-resolved Faraday rotation curves measured with the Gd-BIG sample at the picosecond time scale for different temperatures. This sample was simple for temperature investigation owing to the single RE sublattice of Gd. For temperatures lower than $T = 100$ K, the modulation of the spin-wave mode in the HF sub-THz mode can be observed. The time decay of the HF mode over temperatures ranging from 10 K to 70 K was approximately 20 ps. After 20 ps, the HF precession relaxed, and the spin-wave mode was simultaneously excited, decaying at approximately 100 ps at 12 K. The initial phases of the HF and spin-wave modes heavily depend on the helicity of the circular-pump polarization, which is typical for excitations via the IFE (Fig. 2 in Ref. 27).

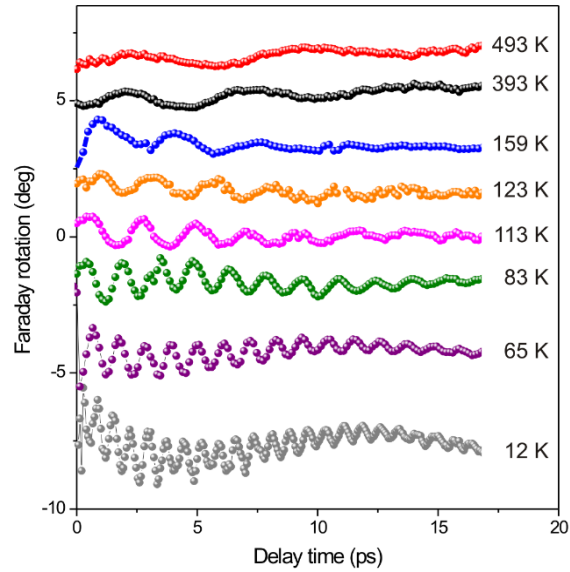

Fig. 9. (Color online) Time-resolved Faraday rotation for different temperatures in Gd-BIG sample.

In general, RE iron garnets can be treated as two-sublattice ferrimagnets, one made up of Fe ions occupying both tetrahedral (d) and octahedral [a] sites ($M_d - M_a = M_{Fe}$) and the other of RE ions occupying dodecahedral {c} sites ($M_c = M_{Gd}$). In our case, total magnetization is given by $M_{Fe} - M_{Gd}$. We obtained the temperature dependence of the Gd and Fe sublattice magnetizations using molecular field theory with a single molecular-field coefficient [27]. The Gd ions exhibit antiferromagnetic exchange coupling with the Fe ions, and the Gd ions tend to align their spins against the net moment of the Fe ions. Because this coupling is relatively weak, the Gd ion ordering is significant only at low temperatures (<60 K), at which the Gd sublattice is much more strongly magnetized than the Fe sublattice [27].

As the temperature decreases below $T_{comp}$, the frequency of the HF mode tends to increase with an increasing $M_s$, which suggests an excitation of the Kaplan–Kittel exchange resonance. In the Gd-BIG sample, this resonance exists because of the exchange interaction between the Gd and Fe sublattices. The magnetizations of these sublattices are noncollinear (inset in Fig. 10) during the entire precession (Fig. 9). Below 60 K, the Fe ions are strongly coupled, and excitations between them are difficult [22]. Thus, in garnets, the Gd–Fe coupling is relatively weak, and the Gd–Gd coupling is even weaker and hence neglected. The magnetization precession in our garnets occurs in a molecular field close to 500 kOe [36,37].

By fitting the time-resolved Faraday rotation data (Fig. 9) with two frequencies, we were able to determine the temperature dependence of both mode frequencies. As the total magnetization increased, the HF mode increased in frequency within the GHz–THz range via exchange resonance excitation (see Fig. 10).

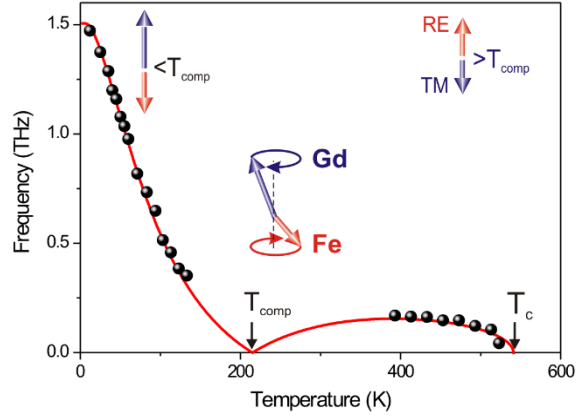

Fig. 10. (Color online) Temperature dependence of the exchange resonance mode (points) for the Gd-BIG sample. The solid line $f_{ex}(T)$ was obtained using Eq. (3).

The temperature dependence of the HF mode provides direct evidence of this resonance. We can calculate the temperature dependence of the HF mode using Eq. (3) for the exchange resonance between the Fe and Gd sublattices. The gyromagnetic ratio $\gamma$ is the same for these sublattices ($\gamma = \gamma_{Gd} = \gamma_{Fe}$) because for Gd ions, $L = 0$. The red solid line in Fig. 10 is calculated using Eq. (3). The experimental data and theoretical calculations show a good agreement.

## 4. Generation and Localization of Magnetization Precession under Excitation of the Exchange Resonance Mode in Magneto-Plasmonic Au/Garnets

Magneto-plasmonics is one of the most promising manipulation methods of spins in the nanometer localization in space [38,39]. In magneto-plasmonic dielectric-based heterostructures, the non-magnetic (noble) metallic layer provides high-quality collective resonances of the free electron gas, whereas magnetic properties are introduced usually by the iron garnets with different substitutions [6]. The latter enables external control of the interface resonances by means of either static magnetic field or the laser-induced effective magnetic field generated via the IFE [15,20,21,26,27]. In Au/garnet heterostructure, two materials with very distinct timescales are combined. The transient electronic response of Au is fast, on the sub-picosecond timescale, whereas garnet lacks free electrons and its relatively slow magneto-optical response is tied to the spin and phonon subsystems.

In the previous part of this review, we demonstrated that a circularly polarized light pulse impulsively excites magnetization precession in GdYb-BIG crystal with HF of exchange resonance mode at about 410 GHz. We assumed that the IFE-induced magnetization dynamics can be enhanced with a localization of the electromagnetic field in such garnet upon the SPP excitation. For that, we studied the SPP-assisted magnetization dynamics in a magneto-plasmonic crystal consisting of 380 μm-thick GdYb-BIG crystal which was complemented with a periodically perforated 50 nm-thick Au overlayer (800 nm period), allowing for the excitation of the SPP mode at the Au/air and Au/garnet interfaces [inset in Fig. 11(a)]. In experiments we used the magneto-optical pump-probe technique with 60 fs temporal resolution. The pump beam wavelength was tuned in the near-IR range [Fig. 11(a) as a purple stripe], while the probe beam wavelength was fixed at 800 nm [red spot on Fig. 11(a)]. The SPP dispersion is shown in Fig. 11(a). By tuning the wavelength and the incidence angle, SPPs can be excited, thus controlling the localization of the hotspot [Fig. 11(a)]. Once the magnetization precession in a garnet is excited via the IFE, the delayed probe beam measures the transient Faraday rotation.

Figure 11 (b) shows the dependences of HF precession of exchange resonance mode at different values of the pump wavelength, and the angle of rotation of the quarter-wave plate for pump pulse. The corresponding values are marked with colored circles in Fig. 11(c). Figure 11(c) shows the dependences of the amplitude and phase of the magnetization precession on the angle of rotation of the quarter-wave plate for two pump wavelengths: 1300 nm (no resonance) and 1380 nm (the vicinity of the SPP resonance at Au/garnet interface). In the non-resonant case (black curves), magnetization precession should be excited in the volume of the ferrite garnet due to the ellipticity of the pump pulse. Accordingly, the precession amplitude will be zero with linear pump polarization, and will increase with a deviation from the zero position of the quarter-wave plate, while the phase will change by $\pi$ for opposite angles of rotation of the plate [Fig. 11(c)]. On the other hand, upon excitation SPP, the precession of magnetization will be excited in the subsurface layer due to the effective plasmon field. Indeed, at a pump wavelength of 1380 nm (red curves), the characteristics of the dependences change [Fig. 11(c)]. This behavior can be explained by the fact that when ellipticity is added to the polarization of the pump pulse, in addition to the magnetization precession in the near-surface layer, the magnetization precession is also excited in the garnet. The phases of these excitations are different; therefore, the resulting angle of the Faraday rotation is determined by their interference.

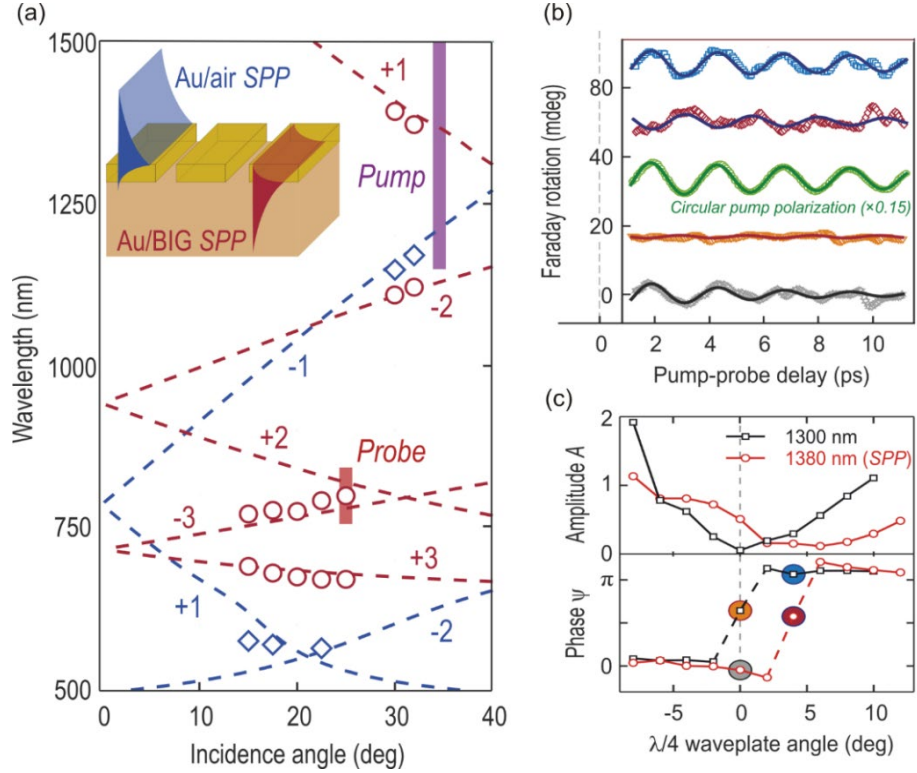

Fig. 11. (Color online) (a) SPP dispersion map of an Au/GdYb-BIG garnet heterostructure. The purple stripe and the red spot show the spectral tunability of the pump and probe pulses, respectively. (b) Time-resolved Faraday rotation with the background removed, together with the fit lines for linear pump polarization at different pump wavelengths. The green data points show the magnetization dynamics excitation with circularly polarized light via IFE. (c) The amplitude and phase of the magnetization precession as a function of the incident pump polarization, where λ/4 wave plate angle of 0° corresponds to the p-polarized excitation. The colored circles correspond to the data shown in (b) with respective colors. The figure is reproduced with permission from Ref. 40 (@2018 American Chemical Society).

To clearly demonstrate the effect of SPP excitation on the amplitude and phase of magnetization precession, we measured their dependence on the pump wavelength. By measuring the variations in the Faraday rotation of the probe pulse with the time delay between the pump and probe, we extracted the amplitude and phase of the exchange resonance mode at different pump-pulse wavelengths. The amplitude spectrum of the exchange resonance mode is shown in Fig. 12 for circular and linear pump polarizations.

Spectral dependences of the circularly polarized pump pulse were monotonic and showed no resonance features. On the contrary, with a linearly polarized pump pulse both in the amplitude and phase of precession, strongly pronounced features were observed near the SPP excitation (pump wavelength of 1400 nm, see Fig. 12). The precession phase at resonance changes by π, which is equivalent to a change in the sign of the amplitude (solid and dashed curves in Fig. 12). We note that in near 1200 nm wavelength, a feature is also observed in the precession phase and amplitude. In this case, the excitation of two different SPPs should occur at different interfaces [40].

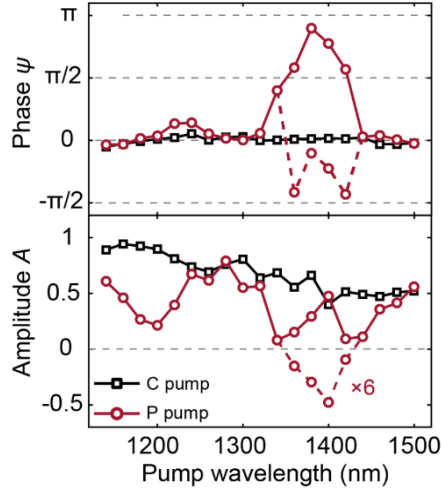

Fig. 12. (Color online) Phase and amplitude of exchange resonance mode versus the wavelength for C-circular (black curve) and P-linear (red curve) pump pulse polarizations. The phase shift of $\pi$ is identical to the amplitude sign changes, as shown by red dashed curves. The figure is reproduced with permission from Ref. 40 (@2018 American Chemical Society).

Here, unlike all the previously demonstrated examples, the SPP affects the sub-THz coherent magnetization dynamics by the spatial confinement of the IFE in a metal–dielectric hybrid structure. As we demonstrated experimentally and through modeling, the SPP electric field rotates in the incidence plane as the SPP propagates along the Au/GdYb-BIG interface and induces a static transversal IFE magnetic field localized on the scale of 100 nm [40]. In the case of circular pump polarization, magnetization precession is excited in depth of a garnet crystal with a thickness of $d_{\text{bulk}}$ = 380 μm (absorption in this wavelength range is low). On the other hand, as shown by numerical calculations [40], with SPP-excitation at the Au/garnet interface, precession should be efficiently excited in a near-surface layer with a depth of about $d_{\text{SPP}} \approx$ 100 nm. Taking into account the difference in precession amplitudes (Faraday rotation $\Delta\theta$) with resonant and non-resonant excitation $\Delta\theta_{\text{SPP}}/\Delta\theta_{\text{bulk}} \approx 0.1$, we obtain following relation between the precession excitation efficiencies with and without SPP excitation: $d_{\text{bulk}} \frac{d_{\text{bulk}}}{d_{\text{SPP}}} \frac{\Delta\theta_{\text{SPP}}}{\Delta\theta_{\text{bulk}}} \approx 400$. Taking advantage of the tunability of the pump wavelength, the excitation efficiency at the SPP resonance was enhanced by two orders of magnitude.

## 5. Exchange Resonance Mode in YIG

Among other rare-earth iron garnets, YIG single crystal possesses superior properties, such as the narrowest FMR linewidth ($\alpha \approx 10^{-5}$) and a Curie temperature $T_c$ of 560 K, and has been regarded as an indispensable magnetic material [41–45]. $Y^{3+}$ ions in the dodecahedral {c} sites are non-magnetic. Therefore, YIG forms a two-sublattice ferrimagnet. Previous studies on YIG have mostly focused on FMR or acoustic magnons in the GHz range. Limited information has been provided about the Kaplan–Kittel exchange resonance mode in the THz frequency range [41]. Recently, several theoretical investigations have revealed full magnon band structures [46–50]. Only inelastic neutron scattering spectroscopy has been experimentally identified for THz magnons in YIG [51–54]. Therefore, the THz-magnon should be investigated using Raman spectroscopy.

Raman spectroscopy was performed in a 180° backscattering geometry with a 785 nm excitation laser. The polarization of the incident and scattered light is denoted as parallel (∥) and crossed (⊥) for linearly polarized configurations and RL, LR, RR, and LL for circularly polarized configurations. The hand of the circularly polarized light (R or L) was defined as the sense of polarization rotation in the sample plane, regardless of the direction of propagation.

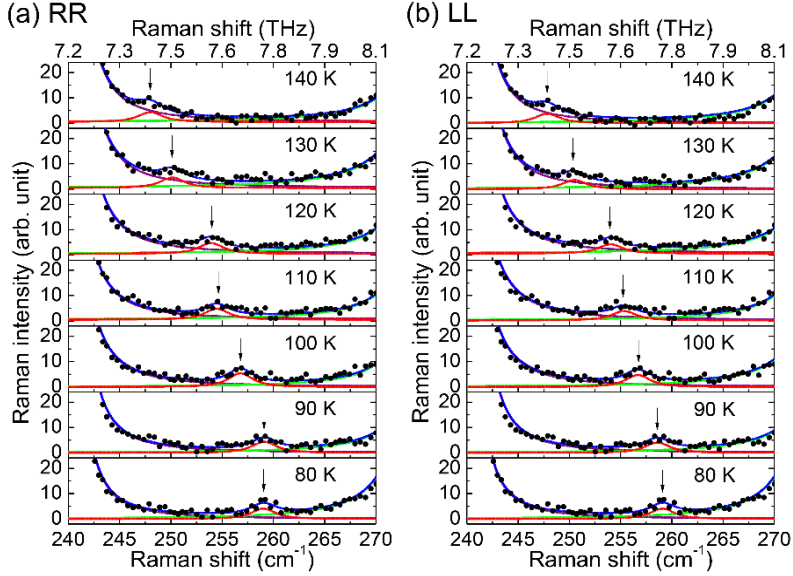

Fig. 13. (Color online) Raman spectra in the 240–270 cm$^{-1}$ range with (a) RR- and (b) LL-polarized configurations in the 80–140 K temperature range in YIG (111). The red, green, and purple curves fit the magnon and two $T_{2g}$ phonons, respectively. The blue curve denotes the sum of the red, green, and purple curves. The figure is reproduced with permission from Ref. 28 (@2020 American Physical Society).

The Raman spectra in the 240–270 cm$^{-1}$ range are depicted in Figs. 13(a) and 13(b) for the RR- and LL-polarized configurations, respectively [28]. In both configurations, a tiny signal was observed and assigned to the resonance exchange mode because of the significant temperature dependence of the frequency shift. At 80 K, the Raman shift was 260 cm$^{-1}$, which corresponds to 7.8 THz.

In Fig. 14(a), the frequency shifts of the THz magnon in the RR- and LL-polarized configurations are plotted as a function of temperature [28] and compared with the results obtained from the neutron scattering measurements [51–54] and simulations [49,50]. The tendency is the same, but a frequency deviation exists between the Raman results and the others. Raman spectroscopy provides a more precise frequency resolution than other techniques. A frequency of 8.0 THz at 4 K was obtained by extrapolating the magnon frequency using the temperature dependence of magnetization [55]. Equation (3) is equivalent to $(12S_d - 8S_a)|J_{ad}| = 10|J_{ad}|$, where a- and d-site spins are $S_a = S_d = 5/2$ and $J_{ad}$ is the exchange constant between the a- and d-site sublattice. This yields $J_{ad} = -38$ K according to $10|J_{ad}| = 8.0$ THz.

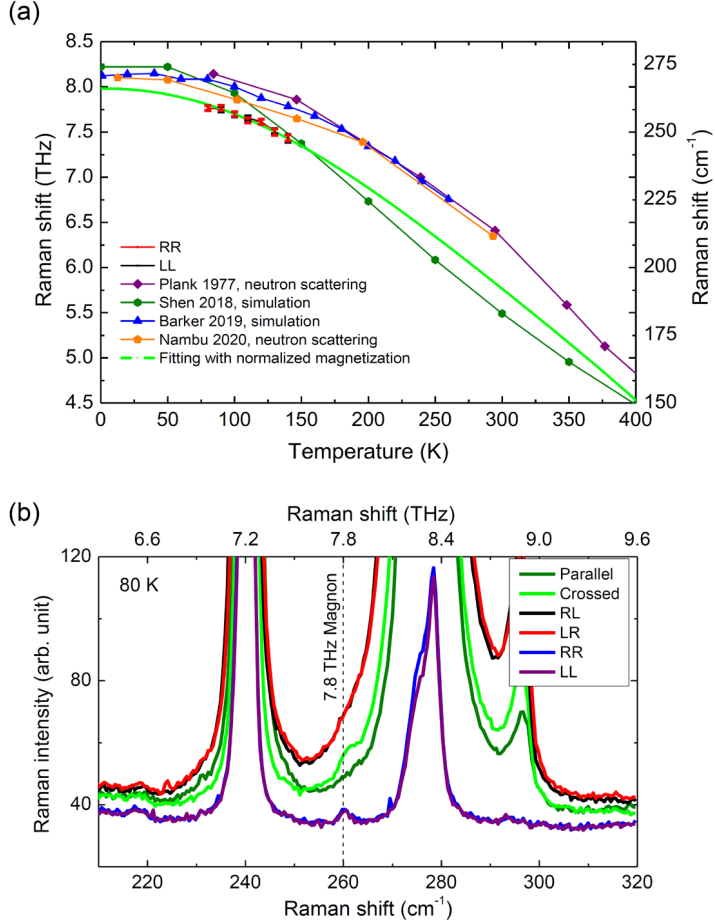

Fig. 14. (Color online) Temperature-dependent frequency shift in the resonance exchange mode (a). The red and black lines with error bars denote the exchange resonance mode observed in our Raman spectra with RR- and LL-polarized configurations, respectively. Selection rule for the resonance exchange mode of YIG observed with crossed-, RR-, and LL-polarized configurations at 260 cm$^{-1}$ (7.8 THz) and 80 K (b). The figure is reproduced with permission from Ref. 28 (@2020 American Physical Society).

The Raman selection rule was investigated at 80 K, as shown in Fig. 14(b). The Raman intensity ratio of the resonance exchange mode was $[I_\parallel : I_\perp : I_{RL} : I_{RR}] = [0 : 1 : 0 : 1]$. This finding is consistent with the magnon Raman tensor in antisymmetric form as

$$R = \begin{pmatrix} 0 & im_z \\ -im_z & 0 \end{pmatrix}, \tag{7}$$

which corresponds to linear (first-order) magnetic excitations [56–60]. The consistency of the exchange resonance mode to the FMR mode excited through light scattering indicates that these modes possess the same symmetry. In the exchange resonance mode, all the atomic spins belonging to the same sublattice simultaneously rotate with their spins always parallel and with the same amplitude, which is analogous to the FMR mode. The spins in the other sublattices behave in the same manner. Therefore, the exchange resonance and FMR modes correspond to the $A_2$ mode [61]. However, unlike the FMR mode, the net magnetization $\mathbf{M}_a + \mathbf{M}_d$ of the exchange resonance mode did not occur. Therefore, the exchange resonance mode is considered to be unobservable via optical methods [62]. Nevertheless, the exchange resonance mode was observed in the Raman method as a linear magnetic excitation, which usually results from the precession of **M**. This is because the Faraday rotation is more sensitive to the a-site Fe sublattice than the d-site Fe sublattice. The exchange resonance mode resulted from the precessional motion of $\mathbf{M}_a$ [29].

## 6. Conclusions

In this review, we have presented the experimental studies and theoretical approaches in RE iron garnets by using ultrafast time-resolved magneto-optical spectroscopy and Raman spectroscopy. Technology for fabricating high-quality garnets at thicknesses of micro- to nano-scales provides an extremely wide variety of compositions. Using magnetic RE-ions substitution for YIG at dodecahedral sites (e.g. Gd, Yb, Lu) in the garnet matrix enables the magnetic ordering and modes of magnetization precession to be varied. We reviewed the non-thermal optically excited precession of magnetization via the IFE and Raman process in RE iron garnet single crystals.

We showed that different frequency modes of the magnetization precession in the GHz and THz ranges strongly depend on the polarization of the pump laser and amplitude of external magnetic field. The low frequency modes in the GHz regime correspond to the spin wave in Bi-substituted iron garnets excitation via the IFE. The HF mode is related to a Kaplan–Kittel exchange resonance mode between magnetic sublattices in ferrimagnetic garnets. The mechanism underlying the exchange interaction plays an important role in understanding the physical origin of magnetism in materials on the atomic level.

A novel approach was using of the garnet crystal with exchange resonance mode at range of sub-THz frequency to study of SPP-driven of magnetization precession with a nanoscale localization. Our experiments and modeling verify the localization of the magnetization dynamics excitation within a 100 nm-thin layer of a transparent dielectric garnet. Taking advantage of the tunability of the pump wavelength, we demonstrated two orders of magnitude enhancement of the excitation efficiency at the SPP resonance. Combining spatial localization with the non-thermal spin–photon coupling, our results open the way toward high-density, low-loss ultrafast optomagnonics.

A wide frequency range of magnetization precession from GHz up to THz and large spatial propagation variations may be useful for wavenumber, magnetic resonance, and dispersion control, as well as offering options for tuning the ultrafast dynamic response with a giant Faraday rotation. Our results demonstrated near-infrared pulse-induced phenomena that are usually observed in the far-infrared range of energies corresponding to atomic interactions in condensed matter. Many related issues remain open for further investigation in optomagnonics.

**Acknowledgment**

A.S. acknowledges support from the grants of the Foundation for Polish Science POIR.04.04.00-00-413C/17-00 and the National Science Centre of Poland (Grant DEC-2017/25/B/ST3/01305). T.S. was financially supported by the Japan Society for the Promotion of Science (JSPS) KAKENHI (Grant Nos. JP19H01828, JP19H05618, and JP19K21854) and Frontier Photonic Sciences Project of National Institutes of Natural Sciences (NINS) (Grant No.

01212002). The authors are particularly grateful to S. Parchenko, A. Chekhov, I. Razdolski, I. Yoshimine, and W.-H. Hsu for their research support.*E-mail: and@uwb.edu.pl

†E-mail: satoh@phys.titech.ac.jp